\documentclass[11pt,a4paper]{article}
\usepackage{jcappub}
  
\usepackage{bm}
\usepackage{epsfig}
\usepackage{citesort}
\usepackage{graphicx}
\usepackage{color}

\begin{document}


\title{Evidence for extra radiation? Profile likelihood versus Bayesian posterior}

\author{Jan Hamann}

\affiliation{Department of Physics and Astronomy\\
 University of Aarhus, DK-8000 Aarhus C, Denmark}

\emailAdd{hamann@phys.au.dk}

\abstract{A number of recent analyses of cosmological data have
  reported hints for the presence of extra radiation beyond the
  standard model expectation.  In order to test the robustness of
  these claims under different methods of constructing parameter
  constraints, we perform a Bayesian posterior-based and a likelihood
  profile-based analysis of current data.  We confirm the presence of
  a slight discrepancy between posterior- and profile-based
  constraints, with the marginalised posterior preferring higher
  values of the effective number of neutrino species $N_{\rm eff}$.
  This can be traced back to a volume effect occurring during the
  marginalisation process, and we demonstrate that the effect is
  related to the fact that cosmic microwave background (CMB) data
  constrain $N_{\rm eff}$ only indirectly via the redshift of
  matter-radiation equality.  Once present CMB data are combined with
  external information about, e.g., the Hubble parameter, the
  difference between the methods becomes small compared to the
  uncertainty of $N_{\rm eff}$.  We conclude that the preference of
  precision cosmological data for excess radiation is ``real'' and not
  an artifact of a specific choice of credible/confidence interval
  construction.}

\maketitle

\section{Introduction}                        \label{sec:introduction}
In the past years, measurements of the temperature and polarisation
anisotropies in the cosmic microwave background have revealed a wealth
of information about the Universe.  One particular quantity that can
be inferred from CMB data is the relativistic energy
density~$\rho_{\rm r}$ around decoupling, typically expressed in terms
of the effective number of massless neutrino degrees of freedom
$N_{\rm eff}$:
\begin{equation}
\rho_{\rm r} = \frac{\pi^2}{15}T_{\gamma}^4 \left( 1 + \alpha
N_{\rm eff}  \right),
\end{equation}
where $T_\gamma = (2.72548 \pm 0.00057)$~K~\cite{Fixsen:2009ug} is the
CMB temperature and $\alpha \equiv \frac{7}{8} \left( \frac{4}{11}
\right)^{4/3}$.  The three standard model neutrino species are
expected to contribute $N_{\rm eff} = 3.046$ effective degrees of
freedom~\cite{Mangano:2005cc}.  Intriguingly however, present
cosmological data show some indication for \mbox{$N_{\rm eff} >
3.046$}~\cite{Hamann:2010bk,Giusarma:2011ex,Keisler:2011aw,Hou:2011ec,Smith:2011es,Hamann:2011ge,Archidiacono:2011gq},
hinting at the possible existence of further light particle species.
These hints are based on a Bayesian statistics analysis of the data
however, and as long as the evidence for $N_{\rm eff} > 3.046$ is
weak, one might also want to consider an alternative approach of
constraining $N_{\rm eff}$.  A profile likelihood analysis for
instance, being prior-independent and parameterisation-invariant,
provides a useful cross-check of these results and is complementary to
the usual Bayesian analysis based on the posterior probability
density~\cite{Hamann:2007pi}.  Using a profile likelihood-based
analysis, it was recently claimed in~\cite{GonzalezMorales:2011ty}
that the hints for $N_{\rm eff} > 3.046$ are merely artifacts of the
Bayesian construction of parameter constraints.  We shall revisit
this claim in the present work.

This paper is organised as follows: we will describe the details of
our analysis in section~\ref{sec:analysis}, present our results in
section~\ref{sec:results} and conclude in
section~\ref{sec:discussion}.

\section{Analysis}                        \label{sec:analysis}

\subsection{Data sets}
For clarity of presentation and given the considerable numerical
effort required to reliably construct the profile likelihood, we will
limit ourselves to two different combinations of data:
\begin{itemize}
\item[1.]{A CMB only set, consisting of
the 7-year Wilkinson Microwave Anisotropy Probe data (WMAP7)
\cite{Komatsu:2010fb} plus the 2008 Atacama Cosmology Telescope (ACT)
data~\cite{Dunkley:2010ge}.  For this data set, the discrepancy between
the Bayesian result of~\cite{Dunkley:2010ge} and the profile
likelihood result reported in~\cite{GonzalezMorales:2011ty} is
particularly large.}
\item[2.]{The same data combined with a constraint on the Hubble
    parameter (HST) derived by Riess {\it et
      al.}~\cite{Riess:2011yx}.}
\end{itemize}

\subsection{Model and priors}
We consider a one-parameter extension of the 6-parameter $\Lambda$CDM
vanilla model, varying also the effective number of massless degrees
of freedom $N_{\rm eff}$ on top of the standard parameters.
Additionally, three parameters describing the foreground contribution
to the small-scale CMB temperature spectrum are required.  The
parameterisation of the vanilla model is not unique, and there are a
number of different choices commonly used in the literature.  Since
the parameterisation implicitly determines the prior probability
distribution, these choices can affect the inference of parameters,
even though the physical models are equivalent.  In this work we
explicitly compare three parameterisation choices: a flat prior on the
Hubble parameter $H_0$, a flat prior on the dark energy density
$\Omega_\Lambda$ and a flat prior on the ratio of sound horizon to
angular diameter distance at decoupling $\theta_{\rm s}$.  We list all
free parameters and their associated prior ranges in
table~\ref{tab:paramsandpriors}.  The primordial Helium fraction is
fixed to $Y_{\rm p} = 0.24$ in order to facilitate comparison with
other authors' results.

\begin{table}[t!]
\caption{Parameters and prior ranges for the cosmological and nuisance
  parameters.  For each individual analysis only one out of the first
  three parameters is used.\label{tab:paramsandpriors}}
\begin{center}
{\footnotesize
\begin{tabular}{|ll|rcl|}
\hline
Parameter   &  Symbol & \multicolumn{3}{|c|}{Prior}   \\
\hline
Hubble parameter & $h$ & 0.4 & $\!\!\!\to\!\!\!$ & 1.0 \\
Dark energy density & $\Omega_{\Lambda}$ & 0 & $\!\!\!\to\!\!\!$ & 1\\
Ratio of sound horizon to angular diameter distance
at decoupling & $\theta_{\rm s}$ & 0.5& $\!\!\!\to\!\!\!$ &10 \\
\hline
Baryon density &$\omega_{\rm b}$ & 0.005& $\!\!\!\to\!\!\!$ &0.1\\
Cold dark matter density & $\omega_{\rm cdm}$ & 0.01& $\!\!\!\to\!\!\!$ &0.99\\
Amplitude of scalar spectrum @ $k=0.05~{\rm Mpc}^{-1}$ &  $\log[10^{10}A_s]$ & 2.7& $\!\!\!\to\!\!\!$& 4\\
Scalar spectral index & $n_{\rm s}$ & 0.5 &$\!\!\!\to\!\!\!$ &1.5\\
Redshift of reionisation & $z_{\rm re}$ & 1& $\!\!\!\to\!\!\!$& 50 \\
\hline
Effective number of massless neutrinos & $N_{\rm eff}$ & 1.5& $\!\!\!\to\!\!\!$ & 10 \\
\hline
Amplitude of Sunyaev-Zel'dovich contribution & $A_{\rm SZ}$ & 0& $\!\!\!\to\!\!\!$&
3\\
Amplitude of clustered point source contribution & $A_{\rm c}$ & 0 &
$\!\!\!\to\!\!\!$ & 20\\
Amplitude of Poisson point source contribution & $A_{\rm P}$ & 0 & $\!\!\!\to\!\!\!$ &
100\\
\hline
\end{tabular}
}
\end{center}
\end{table}

\subsection{Marginalised posterior and profile likelihood}
Given a model with $n$ free parameters, the full information of the
data is contained in the $n$-dimensional likelihood function
$\mathcal{L}$.  If one wants to construct constraints on a single
parameter $\varphi$, the dimensionality obviously needs to be reduced.
Most commonly, this is done in a Bayesian framework, by first
promoting $\mathcal{L}$ to a probability density function (through
multiplication with a prior probability density), and then integrating
(``marginalising'') over the nuisance directions (see
\cite{Hamann:2007pi} for a more detailed discussion), resulting in the
marginalised posterior.  The marginalised posterior can easily be
extracted from Markov chains and has a straightforward interpretation
as the probability density of the true value of $\varphi$, given the
model, data and priors.

Since the choice of priors (or equivalently, the choice of
parameter basis~\cite{Valkenburg:2008cz}) may be somewhat subjective,
one might also want to consider a prior-independent construction, such
as the profile likelihood $\mathcal{L}^{\rm p}$.  Here, instead of
integrating over the nuisance directions, one takes the maximum value
of $\mathcal{L}$ for a fixed value of $\varphi$.  Though the profile
likelihood does not have a formal probabilistic interpretation, it is
often used to construct approximate frequentist confidence intervals
based on the likelihood ratio, by identifying the region for which
$\Delta \chi^2_{\rm eff} \equiv -2 \ln (\mathcal{L}^{\rm
  p}(\varphi_{\rm max}) - \mathcal{L}^{\rm p}(\varphi)) < 1$ with the
68\% confidence interval.  We note that this interval may not have the
desired frequentist coverage properties if the profile likelihood is
not Gaussian~\cite{Porter:1995rz}.

\subsection{Construction of the profile likelihood}
We construct the marginalised posteriors from Markov chains generated
with a modified version of the public Markov chain Monte Carlo
sampler \texttt{CosmoMC}~\cite{Lewis:2002ah}, using a conservative
Gelman-Rubin convergence criterion~\cite{gelru} of $R-1 < 0.01$, and
making sure that the numerical precision settings are sufficient for
the data sets considered.

Na\"{i}vely, one might think that one could use the same chains to
construct the profile likelihood, by binning the data in $N_{\rm eff}$
and identifying the best-fitting point in each bin.  Unfortunately,
this method does not turn out to be suitable for the case at hand.
The reason is that the standard Metropolis-Hastings algorithm samples
the region near the maximum of the posterior very poorly.  In
Appendix~\ref{sec:appendix} we present a rough analytical estimate of
the probability of finding at least one sample of a Markov chain
within a given $\Delta \chi^2_{\rm eff}$ of the best-fit.  As shown in
the bottom panel of figure~\ref{fig:sampling}, for our 10-parameter
model one would need of order $10^5$ independent samples to even have
a $50\%$ chance of the best-fitting sample to lie within 0.5 of the
true best-fit $\chi^2_{\rm eff}$.  This should be compared to the
typically few times $10^4$ correlated samples one usually has in
Markov chains used for parameter estimation.  The problem is
exacerbated by the binning: in particular the estimate of
$\mathcal{L}^{\rm p}$ for the bins in the tails of the marginalised
posterior would be extremely inaccurate.

We therefore employ a different, numerically somewhat more demanding,
construction that avoids under-sampling of the tails and is immune to
biases introduced by a binning procedure.  On a grid of fixed values
of $N_{\rm eff}$, we estimate the respective maxima of the likelihood
by generating Markov chains at temperatures $T \ll 1$, with the
temperature and length of chains chosen such that $\ln
\mathcal{L}^{\rm p}$ is estimated with an accuracy of at least 0.1.
In addition, we determine the global best-fit by letting $N_{\rm eff}$
vary as well.

\section{Results}                        \label{sec:results}

\begin{figure}[t!]
\center
\includegraphics[height=.75\textwidth,angle=270]{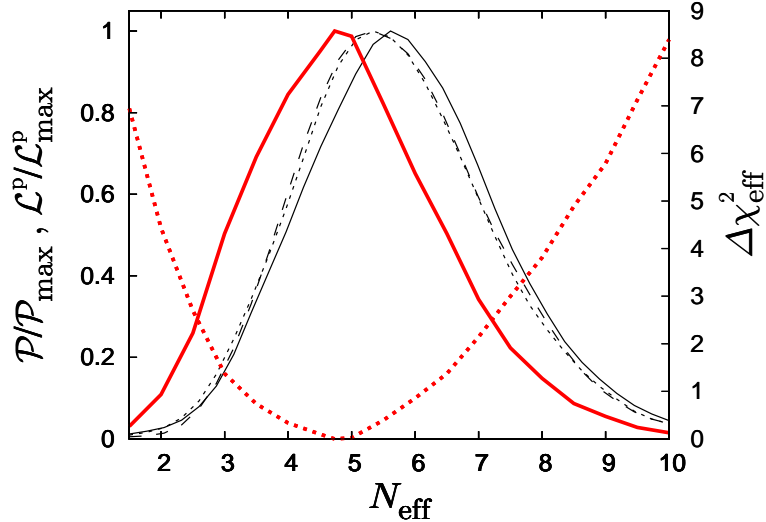}
\caption{Constraints on $N_{\rm eff}$ from WMAP7+ACT data.  Thin black
  lines denote the posterior probability density marginalised over the
  other parameter directions for three different choices of prior
  (solid: $H_0$, dashed: $\theta_{\rm s}$, dotted:
  $\Omega_{\Lambda}$).  The profile likelihood is plotted in thick red
  lines, both in terms of $\mathcal{L}^{\rm p}/\mathcal{L}^{\rm
    p}_{\rm max}$ (solid) and $\Delta \chi^2_{\rm eff}$ (dotted).
  \label{fig:wmapact}}
\end{figure}

In figure~\ref{fig:wmapact} we show the results for WMAP7+ACT data.
Firstly, we note that the posteriors differ very little for the
different priors, indicating a remarkable robustness of the results to
the choice of prior.  Secondly, the profile $\Delta \chi^2_{\rm eff}$
clearly deviates from the parabolic shape one would expect for a
Gaussian profile likelihood, showing an obvious skew towards the
large-$N_{\rm eff}$ side, so we refrain from mapping it to frequentist
confidence limits.  Thirdly, $\mathcal{L}^{\rm p}$ is markedly shifted
(by up to about two thirds of a standard deviation) towards lower
values of $N_{\rm eff}$ compared to the marginalised posteriors.  A
similar tendency was also observed in \cite{Hamann:2007pi}, and, more
recently, in~\cite{GonzalezMorales:2011ty} -- however, their results
for the same data set (both mode and likelihood ratio-based bounds)
differ considerably from ours, possibly due to them attempting to
construct the profile likelihood from Markov chains that were
originally generated for the purpose of Bayesian parameter inference.
For instance, the individual best-fit estimates of the eight $T=1$
WMAP+ACT Markov chains (each containing about $3 \times 10^4$ samples)
we generated for constructing the marginalised posterior display a
considerable spread, with a standard deviation of 0.57 -- indicating
the unreliability of this method.

Is there an explanation for why larger values of $N_{\rm eff}$ have a
high posterior probability despite apparently not fitting the data too
well (and {\it vice versa} for smaller $N_{\rm eff}$)?
In~\cite{GonzalezMorales:2011ty}, it was claimed that the effect, and,
by association, also any possible hints for a deviation of $N_{\rm
  eff}$ from the standard model expectation, is ``driven by prior
effects''.  This is a very generic statement however; it should be
clear that any Bayesian credible intervals are always to some extent
prior-dependent.  We would like to propose a slightly more specific
explanation here, namely that the shift of the marginalised posterior
towards larger $N_{\rm eff}$-enhancement is caused by a volume effect
in the marginalisation process.

Let us, for a moment, imagine the full posterior were Gaussian. In
that case, marginalisation and profiling would lead to the same
result.  Also, for a Gaussian posterior, the variance of the other
parameters' marginalised posteriors on slices of constant $N_{\rm
  eff}$ would not depend on $N_{\rm eff}$.  If, however, these
variances did depend on $N_{\rm eff}$, and happened to be positively
correlated with $N_{\rm eff}$, then at larger (smaller) $N_{\rm eff}$
there would be more (less) volume in the nuisance directions, and the
marginalised posterior would be enhanced (suppressed) compared to the
profile.  We shall see that this is indeed the case here, and there is
in fact a simple physical argument for why it should be so.

\begin{figure}[t!]
\center
\includegraphics[height=.75\textwidth,angle=270]{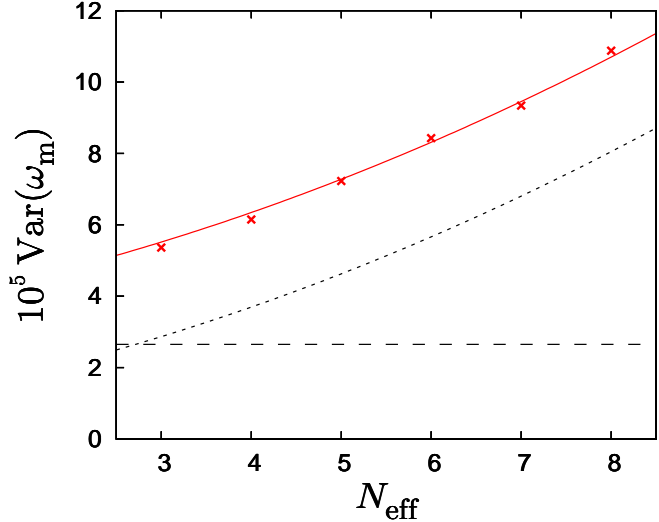}
\caption{Variance of the marginalised posterior probability of
  $\omega_{m}$ on slices of width $\Delta N_{\rm eff} = 1$ as a
  function of $N_{\rm eff}$. The crosses mark the values extracted
  from the Markov chains, the red line is the prediction based on the
  variance of $z_{\rm eq}$, consisting of a constant term induced by the
  bin width (equation~(\ref{eq:binvar}), dashed line) and the intrinsic
  variance of $\omega_{\rm m}$ (equation~(\ref{eq:intvar}), dotted line).
  \label{fig:varomm}}
\end{figure}

As discussed in~\cite{Bashinsky:2003tk,Hou:2011ec}, $N_{\rm eff}$
impacts the CMB power spectra in several ways; most importantly
through the redshift of matter-radiation equality
\begin{equation}
  1 + z_{\rm eq} \equiv \frac{\rho_{\rm m}}{\rho_{\rm r}} =
  \frac{\omega_{\rm m}}{\omega_\gamma} \frac{1}{1+ \alpha \, N_{\rm
  eff}}, 
\end{equation} 
which determines the magnitude of the early integrated Sachs-Wolfe
effect.  It is actually $z_{\rm eq}$ (not $N_{\rm eff}$ or the matter
density~$\omega_{\rm m}$) that is {\it directly} constrained by the
CMB~\cite{Komatsu:2008hk}, and hence essentially uncorrelated with
$\omega_{\rm m}$ and $N_{\rm eff}$.  Ignoring the tiny uncertainty in
the photon energy density $\omega_\gamma$, the variance of
$\omega_{\rm m}$ for fixed $N_{\rm eff}$ is given by
\begin{equation}\label{eq:intvar}
  \left. {\rm Var}(\omega_{\rm m}) \right|_{N_{\rm eff}} \simeq {\rm
    Var}(z_{\rm eq}) \left( \omega_\gamma (1 + \alpha N_{\rm
      eff})\right)^2,
\end{equation}
and thus the posterior becomes wider in the $\omega_{\rm m}$-direction
for larger $N_{\rm eff}$. Since $\omega_{\rm m}$ has degeneracies with
other parameters, such as $H_0$, the widening is propagated to those
directions as well, amplifying the total volume effect.  In
figure~\ref{fig:varomm} we show the $N_{\rm eff}$-dependence of
the posterior's width: using our original Markov chains, we evaluate
the variance of the marginalised posterior of $\omega_{\rm m}$ on
slices of width $\delta N_{\rm eff} = 1$.  This is compared to the
expectation from the measurement of \mbox{$z_{\rm eq} = 3180 \pm
  129$}, which can easily be calculated from the same chains.  The
variance on these slices is composed of two components, the intrinsic
one of equation~(\ref{eq:intvar}), and a constant piece due to the bin
width, given by
\begin{equation}\label{eq:binvar}
  {\rm Var}_{\rm b}(\omega_{\rm m}) = \frac{1}{12} \, \omega_\gamma^2
  z_{\rm eq}^2 \alpha^2 \,  \delta N_{\rm eff}^2.   
\end{equation}
Their sum is found to be in excellent agreement with the variances
from the chains.

\begin{figure}[th]
\center
\includegraphics[height=.75\textwidth,angle=270]{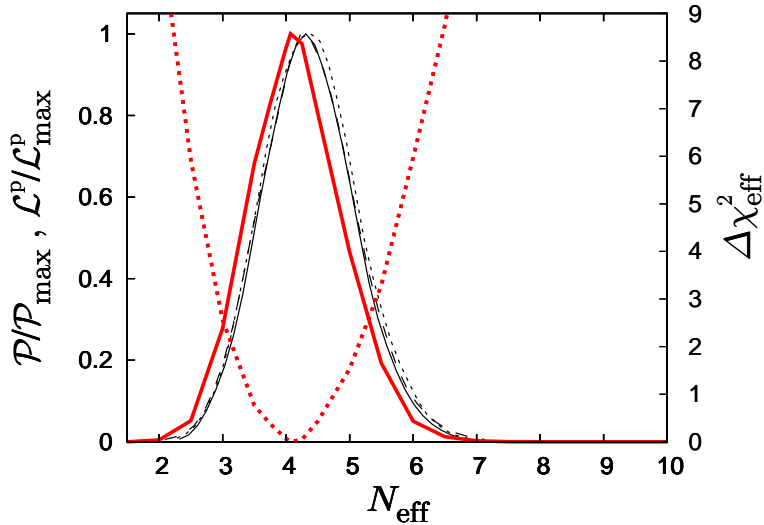}
\caption{Same as figure~\ref{fig:wmapact}, for WMAP7+ACT+HST.\label{fig:wmapacthst}}
\end{figure}

The constraints on $N_{\rm eff}$ can be improved by adding non-CMB
data to break some of the parameter degeneracies, and most of the
recent hints for $N_{\rm eff} > 3.046$ are based on such combinations
of data.  As an example, we add the HST-constraint on $H_0$ here,
which breaks the $N_{\rm eff}$-$H_0$ degeneracy.  Our results for
WMAP7+ACT+HST data are shown in figure~\ref{fig:wmapacthst}.  The
profile likelihood is closer to Gaussian now, and the magnitude of the
volume effect has become much smaller -- $\mathcal{L}^{\rm p}$ is
shifted by roughly 0.2 with respect to the marginalised posteriors.
If we compare this to the posterior standard deviation of $\sim 0.7$,
we see that the volume effect by itself cannot account for the
observed deviation from the standard model expectation.  We summarise
our results in table~\ref{tab:results}.

\begin{table}[th]
  \caption{Summary of constraints on $N_{\rm eff}$ from
    different analysis methods.  For the marginalised posterior we
    list the mean $\langle {N}_{\rm eff} \rangle$, mode
    $\mathcal{P}_{\rm max}$, standard deviation $\sigma_{N_{\rm
        eff}}$, and the minimal 68\%- and 95\%-credible
    intervals~\cite{Hamann:2007pi}.  For the profile likelihood, we
    list the mode $\mathcal{L}^{\rm p}_{\rm max}$ and the intervals in
    which $\Delta \chi_{\rm eff}^2 \leq 1$ and 4,
    respectively. \label{tab:results}} 
\begin{center}
{\footnotesize
\begin{tabular}{|c|ccccc|ccccc|}
\hline
  Analysis &  \multicolumn{5}{|c|}{WMAP7+ACT}  & \multicolumn{5}{|c|}{WMAP7+ACT+HST}   \\
\hline
\hline
 Bayesian &$\langle {N}_{\rm eff} \rangle$ & $\mathcal{P}_{\rm max}$ & $\sigma_{N_{\rm eff}}$ & 68\% MCI & 95\% MCI & $\langle {N}_{\rm eff} \rangle$ & $\mathcal{P}_{\rm max}$ & $\sigma_{N_{\rm eff}}$  & 68\% MCI & 95\% MCI \\
\hline
 $H_0$-prior & 5.78 & 5.68 & 1.45 & 4.18$\to$7.12 & 3.03$\to$8.76 & 4.37 & 4.30 & 0.72 & 3.61$\to$5.03 & 2.96$\to$5.80  \\
 $\theta_{\rm s}$-prior  & 5.69 & 5.20 & 1.44 & 4.02$\to$6.92 & 3.01$\to$8.59 &4.37 & 4.28 & 0.75 & 3.57$\to$5.05 & 2.89$\to$5.86 \\
 $\Omega_\Lambda$-prior  & 5.67 & 5.20 & 1.46 & 4.05$\to$6.98 & 2.90$\to$8.65 & 4.39 & 4.28 & 0.74 & 3.60$\to$5.08 & 2.98$\to$5.89 \\
\hline
\hline
 Profile  & \multicolumn{3}{c}{$\mathcal{L}^{\rm p}_{\rm max}$} & $\Delta \chi_{\rm eff}^2 \leq 1$ & $\Delta \chi_{\rm eff}^2 \leq 4$ & \multicolumn{3}{c}{$\mathcal{L}^{\rm p}_{\rm max}$} &  $\Delta \chi_{\rm eff}^2 \leq 1$ &  $\Delta \chi_{\rm eff}^2 \leq 4$ \\
\hline
 $\mathcal{L}^{\rm p}$ & \multicolumn{3}{c}{4.73} & 3.29$\to$6.14 & 2.12$\to$8.09 &\multicolumn{3}{c}{4.07} & 3.43$\to$4.76 & 2.79$\to$5.50 \\
\hline
\end{tabular}
}
\end{center}
\end{table}

\newpage

\section{Discussion}                        \label{sec:discussion}

We have demonstrated that constraints on the effective number of
neutrino species, inferred from CMB data, can be subject to a slight
discrepancy between the Bayesian marginalised posterior and the
profile likelihood.  This can be attributed to a volume effect
primarily in the $\omega_{\rm m}$ direction, caused by the fact that
the CMB data are directly sensitive mostly to the redshift of
equality, not $N_{\rm eff}$ itself.

Before we come to an interpretation, let us illuminate the statistical
aspect of this result.  Regarded from a sampling theory perspective,
the mode of the full multi-dimensional posterior can be regarded as an
unbiased estimator of the true parameter values (since in the present
problem it coincides by construction with the maximum of the
likelihood).  In the process of marginalisation, this property is lost
-- the most probable value of $N_{\rm eff}$ does not provide the best
possible fit to the data, or, in other words, the mode of
$\mathcal{P}(N_{\rm eff})$ becomes a biased estimator of $N_{\rm eff}$
(see also \cite{Lesgourgues:2006nd} for a discussion,
or~\cite{Stompor:2008sf} for another applied example).  The profile
likelihood on the other hand retains the unbiasedness of the mode
estimator, but, unlike the marginalised posterior, it is not sensitive
to volume effects, and thus does not have a formal statistical
interpretation.

In general it should not come as a surprise that, whenever the full
posterior/likelihood's dimensionality is reduced, loss of information
will be incurred.  Marginalisation and profiling simply preserve
different properties of their related multi-dimensional objects, and
can thus be a good diagnostic of unusual features.  A discrepancy
between the two would point to a deviation from Gaussianity, and, from
a Bayesian perspective, could for instance indicate that a certain
amount of fine-tuning relative to the prior expectation is required in
order to optimise the fit to the data.\footnote{We remark that
  one could, in principle, choose the priors such that the discrepancy
  would vanish (e.g., here, a flat prior on $z_{\rm eq}$ instead of
  $N_{\rm eff}$ might be a good guess).  But with $N_{\rm eff}$
  arguably being a more fundamental quantity than $z_{\rm eq}$, it is
  doubtful whether such a choice could be reasonably justified from a
  theoretical point of view.}

Finally, to evaluate the relevance of this effect, the magnitude of
the bias should be set in relation to the intrinsic width of the
marginal distribution.  For WMAP+ACT data, the difference between
profile and posterior is of order two thirds of a standard deviation,
thus not contributing the dominant -- but certainly a non-negligible
-- part to the indication for a non-standard $N_{\rm eff}$.  With the
addition of HST data, however, the bias is reduced it to less than one
third of a standard deviation, and a similar trend is to be expected
if one added, for instance, large scale structure data, or improved
measurements of the CMB damping tail -- be it existing ones from the
South Pole Telescope~\cite{Keisler:2011aw}, or upcoming ones from
Planck.

We conclude that the recent indication for a deviation of $N_{\rm
  eff}$ from its standard model expectation cannot be accounted for by
this statistical effect alone (though the presence of an additional
statistical bias introduced, e.g., by the modelling of foregrounds,
remains a possibility).

\section*{Acknowledgements}
The author thanks Steen Hannestad and Yvonne Wong for helpful comments
on the manuscript and gratefully acknowledges support from a Feodor
Lynen-fellowship of the Alexander von Humboldt Foundation and the use
of computing resources from the Danish Center for Scientific Computing
(DCSC).


\newpage
\appendix

\section{Profiling with Markov chains}       \label{sec:appendix}
In this section we present an estimate of how well the maximum of
a probability distribution can be determined by using a Markov chain
of length $N$.  

\begin{figure}[th]
\center
\includegraphics[height=.48\textwidth,angle=270]{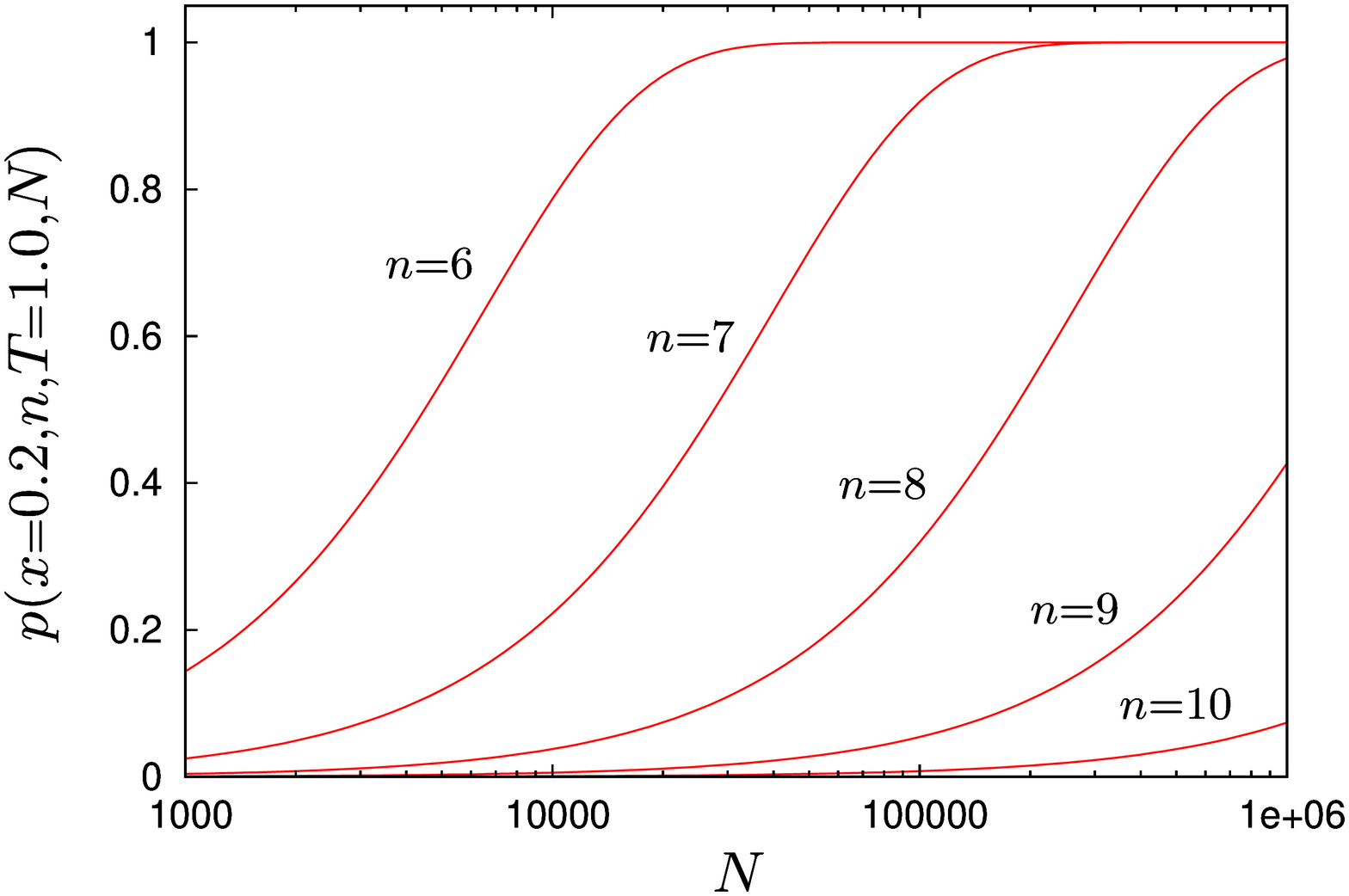}
\includegraphics[height=.48\textwidth,angle=270]{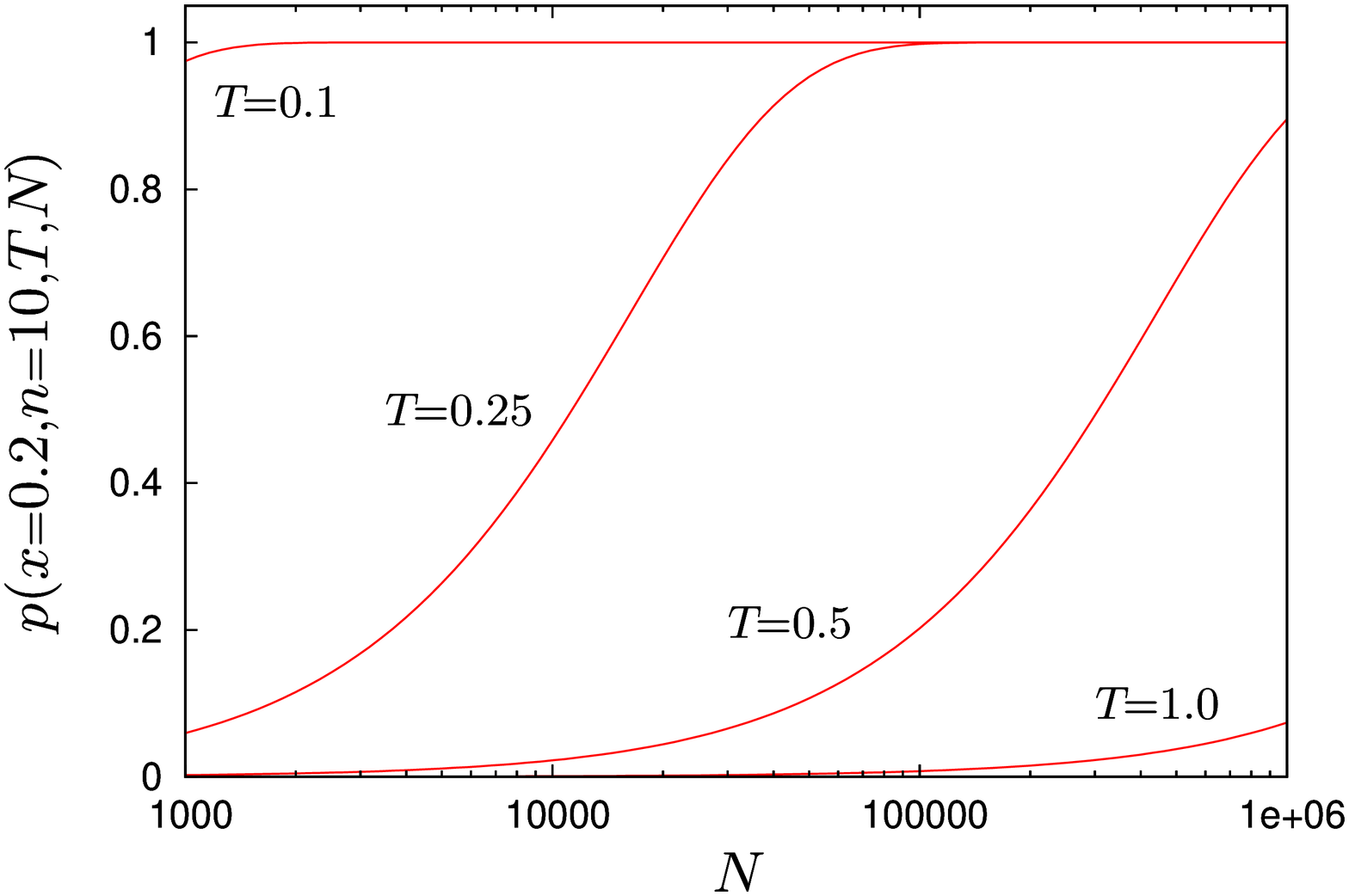}
\includegraphics[height=.48\textwidth,angle=270]{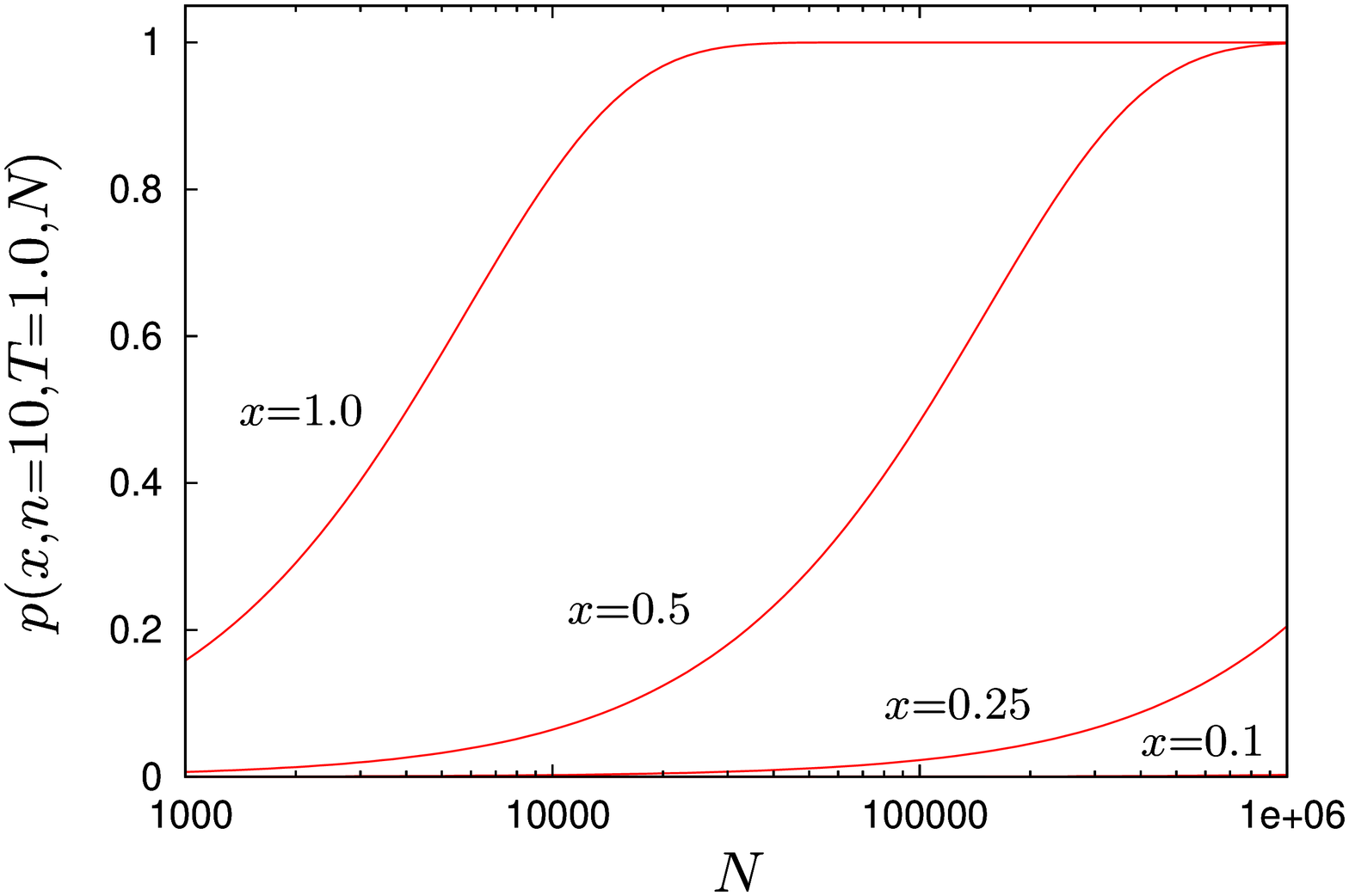}
\caption{Probability of finding at least one sample within $\Delta
  \chi^2_{\rm eff} = x$ of the true maximum of the $n$-dimensional
  Gaussian posterior $\mathcal{P}$, if the Markov chain was generated
  at a temperature $T$ and contains $N$ independent samples. {\it Top
    left:} dependence on $N$ and $n$ for $T=1$ and $x = 0.2$. {\it Top
    right:} dependence on $N$ and $T$ for $n=10$ and $x = 0.2$. {\it
    Bottom:} dependence on $N$ and $x$ for $T=1$ and $n = 10$.
  \label{fig:sampling}}
\end{figure}

Let $\mathcal{P}$ be a probability distribution on an $n$-dimensional
parameter space $\mathfrak{P}$, and $\varphi \in \mathfrak{P}$ be a
point in this parameter space.  We shall make two simplifying
assumptions at this point: first, $\mathcal{P}(\varphi)$ can be
approximated by an $n$-variate Gaussian distribution, and second, all
the samples in the chain are independent.  Without loss of generality
one can then take $\mathcal{P}(\varphi)$ to have unit variance and be
centered around $\varphi_{\rm max} = \vec{0}$.  Define
\begin{equation}
\Delta \chi^2_{\rm eff}(\varphi) \equiv -2 (\ln \mathcal{P}(\varphi_{\rm max})
- \ln \mathcal{P}(\varphi)),
\end{equation}
and the volume fraction $f_x$ of $\mathcal{P}$ for which
$\Delta \chi^2_{\rm eff}(\varphi) < x$,
\begin{equation}
f_x = \int_{V_x} {\rm d} \varphi \; \mathcal{P(\varphi)}, 
\end{equation}
with the volume $V_x$ implicitly given by the condition $\varphi \in
V_x \Leftrightarrow \chi^2_{\rm eff}(\varphi) < x$.  If one expresses
$\varphi$ in spherical coordinates, it can easily be shown that 
\begin{equation}
\label{eq:f1}
  f_x(n) = \int_0^{\sqrt{x}} {\rm d}r \; \frac{1}{\left( \sqrt{2 \pi}
    \right)^{n/2}} \exp \left[ - \frac{r^2}{2} \right] r^{n - 1}
  \frac{\left( 2 \pi \right)^{n/2}}{\Gamma \left( \frac{n}{2} \right)},
\end{equation}
where $\Gamma$ is the Gamma function.  If, instead of sampling from
$\mathcal{P}$, one generates the Markov chain with a temperature
parameter $T$ by sampling from $\mathcal{P}^{1/T}$,
equation~(\ref{eq:f1}) can be generalised to 
\begin{equation}
\label{eq:f2}
f_x(n,T) = \int_0^{\sqrt{x}} {\rm d}r \; \frac{T^{-n/2}}{\left( \sqrt{2 \pi}
  \right)^{n/2}} \left( \exp \left[ - \frac{r^2}{2} \right] \right)^{1/T} r^{n - 1}
\frac{\left( 2 \pi \right)^{n/2}}{\Gamma \left( \frac{n}{2} \right)}.
\end{equation}
If all $N$ samples of the chain are independent, then the probability $\bar{p}$
that none of the points lie within $f_x$ is given by
\begin{equation}
\bar{p}(x,n,T,N) = \left(1 - f_x(n,T) \right)^N. 
\end{equation}
It follows triviallly that the probability of at least one point of
the chain being within \mbox{$\Delta \chi^2_{\rm eff} = x$} of $\chi^2_{\rm
  eff}(\varphi_{\rm max})$ is $p(x,n,T,N) \equiv 1 -
\bar{p}(x,n,T,N)$.  For a few selected slices in $(n,T,N)$-space,
$p(x,n,T,N)$ is plotted in figure~\ref{fig:sampling}.

\end{document}